\documentclass[twocolumn,floatfix,aps,eqsecnum,superscriptaddress]{revtex4-2}
\usepackage{graphicx}
\usepackage{amsmath}
\usepackage{amssymb}
\usepackage{bm}
\usepackage{hyperref}

\usepackage{color}

\begin{document}

\title{Luttinger liquid tensor network:\\
sine versus tangent dispersion of massless Dirac fermions}
\author{V. A. Zakharov}
\affiliation{Instituut-Lorentz, Universiteit Leiden, P.O. Box 9506, 2300 RA Leiden, The Netherlands}
\author{S. Polla}
\affiliation{Instituut-Lorentz, Universiteit Leiden, P.O. Box 9506, 2300 RA Leiden, The Netherlands}
\author{A. Don\'{i}s Vela}
\affiliation{Instituut-Lorentz, Universiteit Leiden, P.O. Box 9506, 2300 RA Leiden, The Netherlands}
\author{P. Emonts}
\affiliation{Instituut-Lorentz, Universiteit Leiden, P.O. Box 9506, 2300 RA Leiden, The Netherlands}
\author{M. J. Pacholski}
\affiliation{Max Planck Institute for the Physics of Complex Systems, N\"{o}thnitzer Strasse 38, 01187 Dresden, Germany}
\author{J. Tworzyd{\l}o}
\affiliation{Faculty of Physics, University of Warsaw, ul.\ Pasteura 5, 02--093 Warszawa, Poland}
\author{C. W. J. Beenakker}
\affiliation{Instituut-Lorentz, Universiteit Leiden, P.O. Box 9506, 2300 RA Leiden, The Netherlands}

\date{July 2024}

\begin{abstract}
To apply the powerful many-body techniques of tensor networks to massless Dirac fermions one wants to discretize the $\bm{p}\cdot\bm{\sigma}$ Hamiltonian and construct a matrix-product-operator (MPO) representation. We compare two alternative discretization schemes, one with a sine dispersion, the other with a tangent dispersion, applied to a one-dimensional Luttinger liquid with Hubbard interaction. Both types of lattice fermions allow for an exact MPO representation of low bond dimension, so they are efficiently computable, but only the tangent dispersion gives a power law decay of the propagator in agreement with the continuum limit: The sine dispersion is gapped by the interactions, evidenced by an exponentially decaying propagator. Our construction of a tensor network with an unpaired Dirac cone works around the fermion-doubling obstruction by exploiting the fact that the \textit{nonlocal} Hamiltonian of tangent fermions permits a \textit{local} generalized eigenproblem.
\end{abstract}
\maketitle

\section{Introduction}
\label{intro}

The linear energy-momentum relation, $E=\pm \hbar v k$, of massless Dirac fermions remains gapless in the presence of disorder, provided that a pair of fundamental symmetries, chiral symmetry and time reversal symmetry, are not both broken \cite{Chi16}. To preserve this so-called topological protection on a lattice one needs to work around the fermion doubling obstruction \cite{Nie81}: if the Brillouin zone contains multiple Dirac cones they can hybridize and open a gap at $E=0$. The nearest-neighbor finite difference discretization suffers from this problem: The resulting sine dispersion, $E= (\hbar v/a) \sin ak$, has a spurious second Dirac cone at the edge $k=\pi/a$ of the Brillouin zone.

It was shown recently \cite{Don22,Bee23} that an alternative discretization of the differential operator, introduced in the 1980's by Stacey \cite{Sta82}, preserves a gapless Dirac cone in a disordered system. The dispersion is a tangent, $E=(2\hbar v/a) \tan (ak/2)$, with a pole rather than a zero at the Brillouin zone edge. No other discretization scheme (staggered fermions, Wilson fermions, {\sc slac} fermions \cite{latticefermions}) has this topological protection. One fundamental consequence is that the Casimir effect for lattice fermions requires the tangent discretization \cite{Bee24}.

All of this is for non-interacting particles. Interacting models of massless Dirac fermions need a lattice formulation for numerical studies \cite{Whi92,Ost95,Sch11,Cir21}, which use methods such as quantum Monte Carlo or DMRG (density matrix renormalization group). The Luttinger liquid with Hubbard interaction, a paradigmatic non-Fermi liquid \cite{Gia03,Giu08}, can be solved analytically in the continuum via bosonization \cite{Lut63,Hal81}, providing a testing ground for lattice calculations. Such a test was reported for quantum Monte Carlo in Ref.\ \onlinecite{Zak24}. Here we consider the DMRG implementation.

The two techniques require a different approach, each with its own challenges. For quantum Monte Carlo the discretization is at the level of the Lagrangian, and the challenge is to ensure a positive action determinant (avoiding the so-called sign problem). For DMRG the discretization involves the representation of the second quantized Hamiltonian by a tensor network \cite{Ver04,Zwo04}: a product of matrices of operators acting locally on each site. The challenge is to ensure that the rank of each matrix (the bond dimension) is small and does not grow with the number of sites. 

 Tangent fermions have a hidden locality originating from the fact that --- although the tangent discretization produces a Hamiltonian with a highly non-local, non-decaying, coupling of distant sites \cite{Sta82} --- the ground state can be obtained from a \textit{local} generalized eigenproblem \cite{Pac21}. Our key finding is that this allows for an exact matrix-product-operator (MPO) representation of low bond dimension. In an independent study \cite{Hae24}, Haegeman \textit{et al.} reached the same conclusion.

In what follows we will compare the sine and tangent discretizations of the Luttinger Hamiltonian, and test the correlators against the continuum results. We first construct the MPO explicitly in Sec.\ \ref{sec_tensornetwork}. The correlators are calculated via the DMRG approach and compared with bosonization in Sec.\ \ref{sec_DMRG}. We conclude in Sec.\ \ref{sec_conclude}. Appendix  \ref{app_GEV} contains the connection between a local generalized eigenproblem and a scale-independent MPO.

\section{Matrix product operator}
\label{sec_tensornetwork}

The starting point of a tensor network DMRG calculation \cite{Sch11} is the representation of the Hamiltonian by a matrix product operator (MPO), to ensure that the variational ground state energy can be computed efficiently for a matrix product state. 

In this section we construct the MPO representation of the one-dimensional (1D) Dirac Hamiltonian
\begin{equation}
H=-i\hbar v\begin{pmatrix}
\partial/\partial x&0\\
0&-\partial/\partial x
\end{pmatrix},
\end{equation}
discretized on a lattice. (The matrix structure refers to the spin degree of freedom.) Once we have done that we will compute the correlators via DMRG in the presence of a Hubbard interaction (Luttinger model).

\subsection{Free fermions}

Consider noninteracting, spinless chiral fermions on a chain of $N$ sites (unit spacing), with hopping matrix elements $t_{nm}$ ($n>m\geq 1$). (We will include the spin degree of freedom and the electron-electron interaction later on.) For an infinite translationally invariant lattice, $t_{nm}=t(n-m)$ is a Fourier coefficient of the dispersion relation,
\begin{equation}
E(k)=2\operatorname{Re}\sum_{n=1}^\infty t(n)e^{ink}.
\end{equation}

The second quantized Hamiltonian
\begin{equation}
H=\sum_{n>m=1}^N \bigl(t_{nm}^{\vphantom{\dagger}}c_n^\dagger c_m^{\vphantom{\dagger}}+t_{nm}^{\ast}c_m^\dagger c_n^{\vphantom{\dagger}}\bigr)\end{equation}
can be rewritten as a product of matrices $M^{(n)}$ that act only on site $n$, but the dimension of each matrix (the bond dimension) will typically grow linearly with $N$.

An exact MPO representation with scale-independent bond dimension is possible in two cases \cite{McC08,Cro08,Pri10,Fro10}: for a short-range hopping ($t_{nm}\equiv 0$ for $n-m>r$) and for a long-range hopping with a polynomial-times-exponential distance dependence:
\begin{equation}
t_{nm}=2t_0 e^{i\phi} (n-m)^p e^{\beta(n-m)},\;\;\beta\in\mathbb{C},\;\;p\in\mathbb{N},\label{tnmdef}
\end{equation}
and linear combinations of this functional form. While the exponent $\beta=\beta_1+i\beta_2$ can be an arbitrary complex number, the power $p$ must be a non-negative integer \cite{Fro10}. A decaying $t_{nm}\propto 1/(n-m)^p$ does not qualify.

The sine dispersion corresponds to a short-range, nearest-neighbor hopping,
\begin{equation}
t_{nm}=(t_0/2i)\delta_{n-m,1}\Leftrightarrow E(k)=t_0\sin k.\label{tnmdefsine}
\end{equation}
The MPO Hamiltonian has bond dimension 4,
\begin{subequations}
\label{MPOsine}
\begin{align}
&H_{\rm sine}=\tfrac{1}{2}it_0[M^{(1)}M^{(2)}\cdots M^{(N)}]_{1,4},\\
&M^{(n)}=\begin{pmatrix}
1&c_n&c_n^\dagger&0\\
0&0&0&c_n^\dagger\\
0&0&0&c_n\\
0&0&0&1
\end{pmatrix}.
\end{align}
\end{subequations}

A no-go theorem \cite{Nie81} forbids short-range hopping if one wishes to avoid fermion doubling and preserve chiral symmetry. If we also require a scale-independent bond dimension we need the hopping \eqref{tnmdef}. In the simplest case $p=0$ of a purely exponential distance dependence \cite{note1}, one has the dispersion
\begin{equation}
E(k)=2t_0\frac{e^{\beta_1} \cos\phi-\cos (\beta_2+k+\phi)}{\cos (\beta_2+k)-\cosh \beta_1}.
\end{equation}
This should be a continuous function in the interval $(-\pi,\pi)$, crossing $E=0$ at $k=0$ but not at any other point in this interval. The only parameter choice consistent with these requirements is $\phi=\pi/2$, $\beta_1=0$, $\beta_2=\pi$, when
\begin{equation}
t_{nm}=2it_0 (-1)^{n-m}\Leftrightarrow E(k)=2t_0\tan(k/2).\label{tnmdeftangent}
\end{equation}
This is Stacey's tangent dispersion \cite{Sta82,note0}.

The corresponding MPO Hamiltonian is
\begin{subequations}
\label{MPOtangent}
\begin{align}
&H_{\rm tangent}=2it_0[M^{(1)}M^{(2)}\cdots M^{(N)}]_{1,4},\\
&M^{(n)}=\begin{pmatrix}
1&c_n&c_n^\dagger&0\\
0&-1&0&c_n^\dagger\\
0&0&-1&c_n\\
0&0&0&1
\end{pmatrix},
\end{align}
\end{subequations}
again with bond dimension 4, differing from the sine MPO \eqref{MPOsine} by the $-1$'s on the diagonal.

\subsection{Helical Luttinger liquid}

We next include the spin degree of freedom and consider helical instead of chiral fermions,
\begin{equation}
H=\sum_{n>m=1}^N \left [t_{nm}\bigl(c_{n\uparrow}^\dagger c_{m\uparrow}^{\vphantom{\dagger}}-c_{n\downarrow}^\dagger c_{m\downarrow}^{\vphantom{\dagger}}\bigr)+\text{H.c.}\right]+\sum_{n=1}^N U_n.\label{HHubbard}
\end{equation}
(H.c. denotes the Hermitian conjugate.) We have added an on-site Hubbard interaction,
\begin{equation}
U_i=U(n_{i\uparrow}-\tfrac{1}{2})(n_{i\downarrow}-\tfrac{1}{2}),\;\;n_{i\sigma}=c_{i\sigma}^\dagger c_{i\sigma}^{\vphantom{\dagger}}.
\end{equation}

The MPO representation for the tangent discretization \eqref{tnmdeftangent} is
\begin{subequations}
\label{MPOtangenthelical}
\begin{align}
&H_{\rm tangent}=2it_0[M^{(1)}M^{(2)}\cdots M^{(N)}]_{1,6},\\
&M^{(n)}=\begin{pmatrix}
1&c_{n\uparrow}&c_{n\uparrow}^\dagger&c_{n\downarrow}&c_{n\downarrow}^\dagger&(2it_0)^{-1}U_n\\
0&-1&0&0&0&c_{n\uparrow}^\dagger\\
0&0&-1&0&0&c_{n\uparrow}\\
0&0&0&-1&0&-c_{n\downarrow}^\dagger\\
0&0&0&0&-1&-c_{n\downarrow}\\
0&0&0&0&0&1
\end{pmatrix},
\end{align}
\end{subequations}
with bond dimension 6. For the sine discretization the $-1$'s on the diagonal are replaced by $0$'s,
\begin{subequations}
\label{MPOsinehelical}
\begin{align}
&H_{\rm sine}=\tfrac{1}{2}it_0[M^{(1)}M^{(2)}\cdots M^{(N)}]_{1,6},\\
&M^{(n)}=\begin{pmatrix}
1&c_{n\uparrow}&c_{n\uparrow}^\dagger&c_{n\downarrow}&c_{n\downarrow}^\dagger&(\tfrac{1}{2}it_0)^{-1}U_n\\
0&0&0&0&0&c_{n\uparrow}^\dagger\\
0&0&0&0&0&c_{n\uparrow}\\
0&0&0&0&0&-c_{n\downarrow}^\dagger\\
0&0&0&0&0&-c_{n\downarrow}\\
0&0&0&0&0&1
\end{pmatrix}.
\end{align}
\end{subequations}

To deal with fermionic statistics, we apply the Jordan-Wigner transformation to the MPOs (see App.~\ref{app_JW}).

The MPOs written down so far refer to an open chain of $N$ sites. To minimize finite-size effects periodic boundary conditions are preferrable: the chain is wrapped around a circle, and sites $n$ and $n+N$ are identified. A translationally invariant hopping, $t_{nm}=t(n-m)$, then requires
\begin{equation}
	t(N-n)=t(n)^\ast,\;\;1\leq n\leq N-1.
\end{equation}

For $N$ odd the Hamiltonian in the tangent discretization \eqref{tnmdeftangent} satisfies this condition without further modification: because of the all-to-all hopping a closing of the chain on a circle makes no difference. (For $N$ even one would have antiperiodic boundary conditions \cite{note2}.) We can therefore still use the MPO \eqref{MPOtangenthelical}.

The sine discretization \eqref{tnmdefsine} requires an additional hopping term between sites 1 and $N$. We construct this MPO explicitly in App.~\ref{app_PBC_sine}.

\section{Correlators}
\label{sec_DMRG}

\subsection{Free fermions}

The propagator
\begin{equation}
C_\sigma(x,x')=\langle c_\sigma^\dagger(x)c_\sigma(x')\rangle,\;\;\sigma\in\{\uparrow,\downarrow\}\leftrightarrow\{ 1,-1\},
\end{equation}
of a non-interacting 1D Dirac fermion with dispersion $E(k)=\pm\hbar vk$ can be readily evaluated:
\begin{align}
C_\sigma(x,x')={}&\frac{1}{Z}\operatorname{Tr}e^{-\beta H}c_\sigma^\dagger(x)c_\sigma(x')\nonumber\\
={}&\int_{-\infty}^\infty\frac{dk}{2\pi}\frac{e^{ik(x-x')}}{1+e^{\beta E(k)}}\nonumber\\
={}&\frac{\sigma \hbar v}{2i\beta\sinh[\pi(\hbar v/\beta)(x-x')]},
\end{align}
for $x\neq x'$, with $Z=\operatorname{Tr}e^{-\beta H}$ the partition function at inverse temperature $\beta=1/k_{\rm B}T$. This reduces to
\begin{equation}
\lim_{\beta\rightarrow\infty}C_\sigma(x,x')=\frac{\sigma}{2\pi i(x-x')}\label{Cbetainfinity}
\end{equation}
in the zero-temperature limit.

On a lattice ($x/a=n\in\mathbb{Z}$, $c_\sigma(x=na)\equiv c_\sigma(n)$) the integration range of $k$ is restricted to the interval $(-\pi/a,\pi/a)$. In the zero-temperature limit, with $\sigma E(k)<0$ for $-\pi/a<k<0$, one then finds
\begin{align}
C_\sigma(n,m)={}&\sigma\int_{-\pi/a}^0 \frac{dk}{2\pi} e^{ika(n-m)}\nonumber\\
={}&\begin{cases}
\frac{2 \sigma}{2\pi  i a(n-m)}&\text{if}\;\;n-m\;\;\text{is odd},\\
0&\text{if}\;\;n-m\;\;\text{is even},
\end{cases}
\end{align}
irrespective of the functional form of the dispersion relation $E(k)$. The continuum result \eqref{Cbetainfinity} is only recovered if one averages over even and odd lattice sites.

The even-odd oscillation also appears in the transverse spin correlator,
\begin{equation}
R(x,x')=\tfrac{1}{4}\langle \bm{c}^\dagger(x){\sigma}_x\bm{c}(x) \;\;\bm{c}^\dagger(x'){\sigma}_x\bm{c}(x')\rangle,
\end{equation}
defined in terms of the spinor $\bm{c}=(c_\uparrow,c_\downarrow)$ and Pauli matrix ${\sigma}_x$. 

For free fermions Wick's theorem gives
\begin{equation}
R(x,x')=-\frac{1}{4}\biggl(C_\uparrow(x,x')C_\downarrow(x',x)+C_\downarrow(x,x')C_\uparrow(x',x)\biggr),
\end{equation}
which at zero temperature results in
\begin{equation}
R(x,x')=\frac{1}{2}\,[2\pi(x-x')]^{-2},\label{Rxcont}
\end{equation}
in the continuum and
\begin{equation}
R(n,m)=\begin{cases}
2[2\pi a(n-m)]^{-2}&\text{if}\;\;n-m\;\;\text{is odd},\\
0&\text{if}\;\;n-m\;\;\text{is even}.
\end{cases}
\end{equation}

The even-odd oscillation \cite{Wan23} can be removed in a path integral formulation, by discretizing the Lagrangian in both space and (imaginary) time \cite{Zak24}, but in the Hamiltonian formulation considered here it is unavoidable. In what follows we will consider smoothed lattice correlators, defined by averaging the fermionic operators $c_\sigma(n)$ over nearby lattice sites. The precise form of the smoothing profile will not matter in the continuum limit $a\rightarrow 0$, we take the simple form
\begin{equation}
\bar{c}_{n\sigma}=\tfrac{1}{2}c_{n\sigma}+\tfrac{1}{2}c_{n+1\sigma},\label{barcdef}
\end{equation} 
so an equal-weight average over adjacent sites. The smoothed correlators are then defined by
\begin{subequations}
\begin{align}
&\bar{C}_\sigma(n,m)=\langle\bar{c}_{n\sigma}^\dagger\bar{c}_{m\sigma}\rangle,\label{barCdef}\\
&\bar{R}(n,m)=\tfrac{1}{4}\langle \bm{\bar{c}}^\dagger_n{\sigma}_x\bm{\bar{c}}_n \;\;\bm{\bar{c}}^\dagger_m{\sigma}_x\bm{\bar{c}}_m\rangle.\label{barRdef}
\end{align}
\end{subequations}

\subsection{DMRG calculation with Hubbard interaction}

We represent the ground state wave function $\Psi$ of the Luttinger liquid Hamiltonian \eqref{HHubbard} by a matrix product state (MPS) and carry out the tensor network DMRG algorithm \cite{Sch11} to variationally minimize $\langle\Psi|H|\Psi\rangle/\langle\Psi|\Psi\rangle$. (We used the TeNPy Library \cite{tenpy} for these calculations.) 
We compare the results for tangent and sine discretization.
The MPOs for both are exact with small bond dimension (given explicitly in App.~\ref{app_JW} and App.~\ref{app_PBC_sine}). The bond dimension $\chi$ of the MPS is increased until convergence is reached (see App.\ \ref{app_DMRG}).

The Luttinger liquid is simulated at zero temperature ($\beta\rightarrow\infty$) and at fixed particle number ${\cal N}={\cal N}_\uparrow+{\cal N}_\downarrow$ (canonical ensemble). We take $N=51$ an odd integer, with periodic boundary conditions for the MPO. The periodicity of the MPS is not prescribed \textit{a priori}, to simplify the DMRG code. By setting ${\cal N}_\uparrow=(N+1)/2$ and ${\cal N}_\downarrow=(N-1)/2$ we model a half-filled band.

The bosonization theory of an infinite Luttinger liquid gives a power law decay of the zero-temperature, zero-chemical-potential correlators \cite{Gia03},
\begin{subequations}
\begin{align}
&C_\sigma(x,x')\propto |x-x'|^{-(1/2)(K+1/K)},\\
&R(x,x')\propto |x-x'|^{-2K},\\
&K=\sqrt{(1-\kappa)/(1 +\kappa)},\;\;\kappa=\frac{U}{2\pi t_0}\in(-1,1).
\end{align}\label{powerlaw}
\end{subequations}
For repulsive interactions, $U>0\Rightarrow K<1$, the transverse spin correlator $R$ decays more slowly than the $1/x^2$ decay expected from a Fermi liquid.

\begin{figure}[tb]
\centerline{\includegraphics[width=0.9\linewidth]{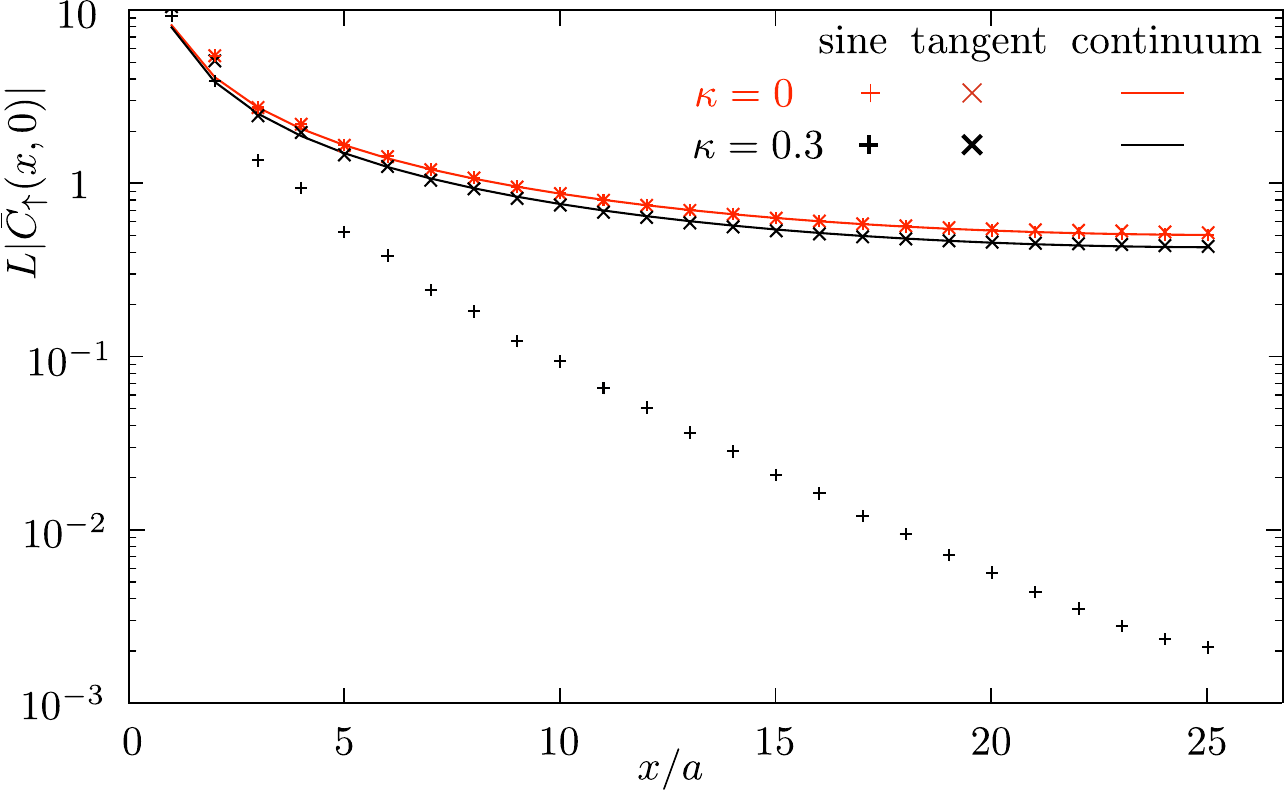}}
\caption{Data points: absolute value of the propagator $\bar{C}_\sigma(n,m)$, defined in Eq.\ \eqref{barCdef} for $\sigma=\uparrow$, $n=x/a$ and $m=0$, calculated in the tensor network of $L/a=51$ sites and bond dimension $\chi=4096$ of the matrix-product state. Results are shown for the tangent and sine discretization of the Luttinger Hamiltonian, for free fermions and for a repulsive Hubbard interaction of strength $\kappa=U/2\pi t_0=0.3$. The curves are the analytical results in the continuum.}
\label{fig_Cplot}
\end{figure}
\begin{figure}[tb]
\centerline{\includegraphics[width=0.9\linewidth]{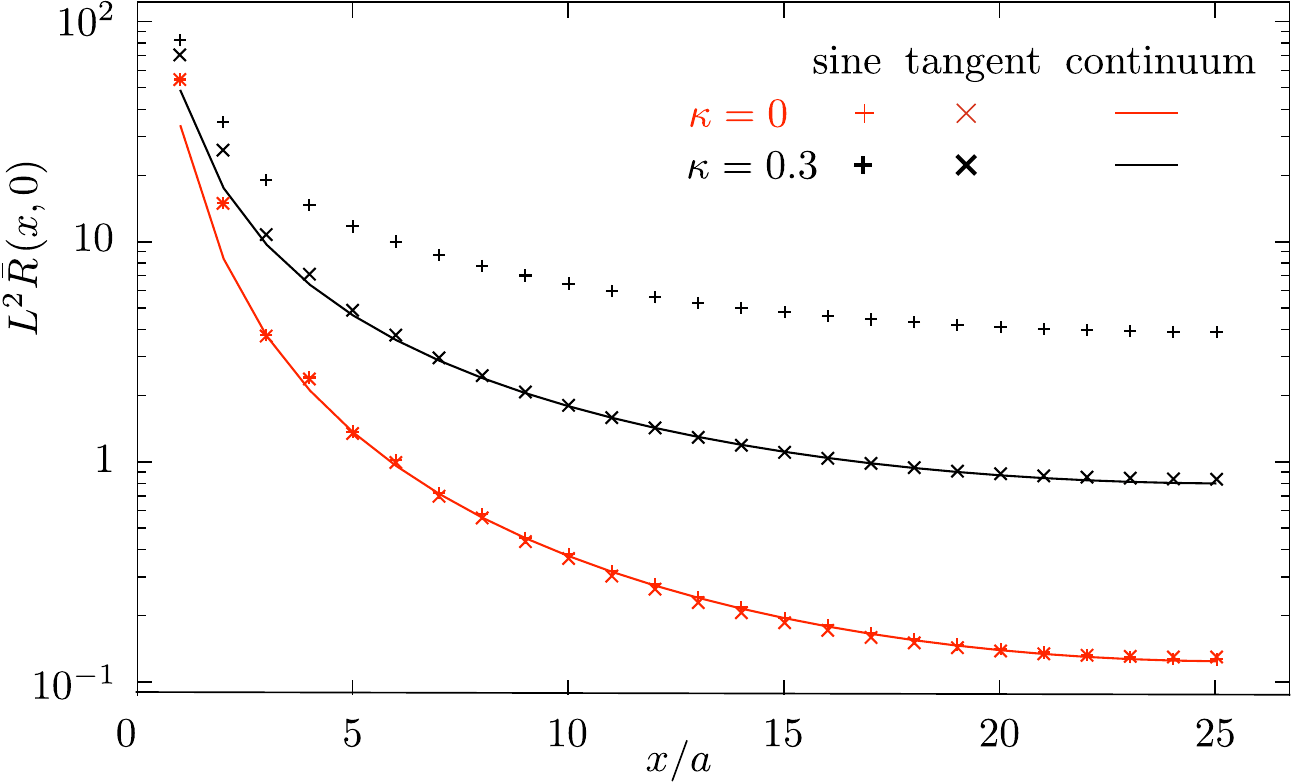}}
\caption{Same as Fig.\ \ref{fig_Cplot}, but now for the transverse spin correlator $\bar{R}(n,m)$ defined in Eq.\ \eqref{barRdef}.}
\label{fig_Rxplot}
\end{figure}

The numerical results are shown in Figs.\ \ref{fig_Cplot} and \ref{fig_Rxplot} (data points). The curves are the continuum bosonization formulas (including finite-size corrections, see App.\ \ref{app_bosonization}). The lattice calculations with the tangent dispersion (crosses) agree nicely with the continuum formulas, without any adjustable parameter. The sine dispersion (plusses), in contrast, only agrees for free fermions. With interactions the sine dispersion gives an exponential decay of the propagator, indicative of the opening of an excitation gap.

\section{Conclusion}
\label{sec_conclude}

We have constructed a Hamiltonian-based tensor network formulation of a Luttinger liquid on a 1D lattice, complementing the Lagrangian-based path integral formulation of Ref.\ \onlinecite{Zak24}. The key step is the Hermitian discretization of the momentum operator $-i\hbar d/d x$ in a way that preserves the fundamental symmetries (chiral symmetry and time reversal symmetry) of massless Dirac fermions. We have compared two discretizations, both allowing for a tensor network of low, scale-independent bond dimension. In this concluding section we also discuss a third.

\begin{figure}[tb]
\centerline{\includegraphics[width=1\linewidth]{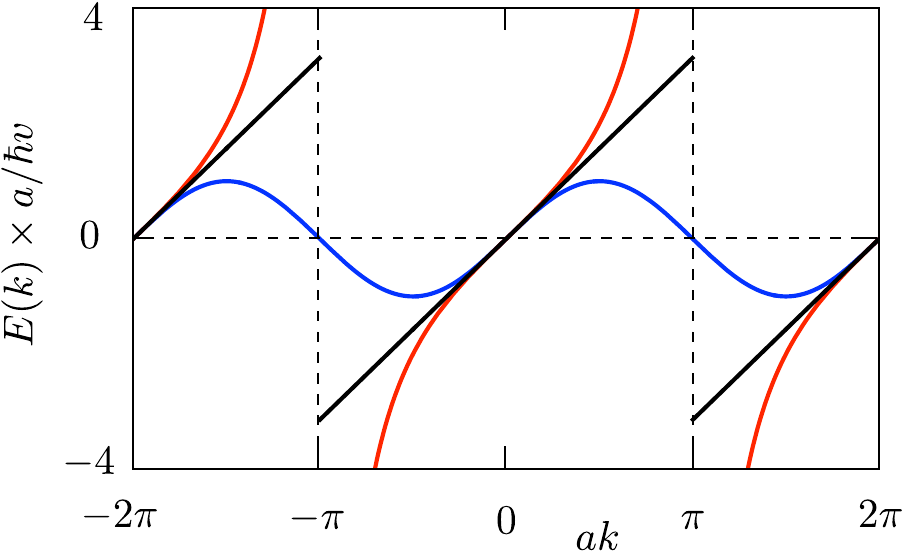}}
\caption{The three ways to discretize the derivative operator in Eq.\ \eqref{dfdxthree} produce three different dispersion relations: sine (blue), tangent (red), and sawtooth (black). The energy-momentum relation of a chiral fermion is obtained from the discretized derivative by substituting $f(x+na)=e^{inka}f(x)$ and equating $-i\hbar vdf/dx=Ef$. The tangent and sawtooth dispersions are discontinuous at the Brillouin zone boundaries ($k=\pm\pi/a$), where the sine dispersion has a second root (fermion doubling). The three dispersion relations coincide near $k=0$, so the corresponding discretized derivatives are equivalent if applied to functions that vary smoothly on the scale of the lattice spacing. }
\label{fig_dispersion}
\end{figure}

The three discretizations of the differential operator on a 1D lattice (unit lattice constant $a$) are the following:
\begin{subequations}
\label{dfdxthree}
\begin{align}
\frac{df}{dx}\mapsto{}&\tfrac{1}{2}[f(x+1)-f(x-1)]\;\;\text{(sine dispersion)},\\
\frac{df}{dx}\mapsto{}& 2\sum_{n=1}^\infty(-1)^n[f(x-n)-f(x+n)]\nonumber\\
&\text{(tangent dispersion)},\\
\frac{df}{dx}\mapsto{}& \sum_{n=1}^\infty(-1)^n \frac{1}{n}[f(x-n)-f(x+n)]\nonumber\\
&\text{(sawtooth dispersion)}.
\end{align}
\end{subequations}
The corresponding dispersion relations are shown in Fig.\ \ref{fig_dispersion}. The energy-momentum relation is a \textit{sine} for the nearest-neighbor difference and a \textit{tangent} for the long-range Stacey derivative \cite{Sta82}. The third dispersion is a (piecewise linear) \textit{sawtooth}, produced by a nonlocal discretization known as the {\sc slac} derivative in the particle physics literature \cite{Dre76}.

The sine dispersion suffers from fermion doubling \cite{latticefermions} --- a second species of low-energy excitations appears at the Brillouin zone boundary. The tangent and sawtooth dispersion describe an unpaired chiral fermion, they rely on nonlocality to work around the theorem \cite{Nie81} that requires chiral fermions to come in pairs in any local theory on a lattice. 

Both the Stacey derivative and the {\sc slac} derivative couple arbitarily distant sites $n,m$, the former $\propto (-1)^{n-m}$ and the latter $\propto (-1)^{n-m}\times(n-m)^{-1}$. From the perspective of a tensor network there is an essential difference between the two: Because the MPO condition \eqref{tnmdef} allows for an exponential distance dependence but excludes a coupling that decays as a power law with distance, only the tangent dispersion has an exact MPO representation with scale-independent bond dimension --- the sawtooth dispersion does not. Tangent fermions have a hidden locality, their spectrum is governed by a \textit{local} generalized eigenproblem \cite{Pac21}, which is at the origin of the efficient tensor network (see App.\ \ref{app_GEV}).

The method we developed here enables simulations of systems with various filling factors and scalar potentials, including those with disorder. By focusing on the impurity-free Luttinger liquid we could in this work test the numerical approach against analytical formulas. The close agreement gives us confidence that tangent fermion DMRG is a reliable method, which at least in 1D is highly efficient.

The next step is to apply it to problems where no analytics exists, condensed matter and particle physics provide a variety of such problems. One class of applications is the stability of gapless chiral modes to the combination of disorder and interactions. Existing DMRG studies \cite{Zen22} work around the fermion doubling obstruction by studying a strip geometry with two edges --- tangent fermions would allow for a single-edge implementation.

For such applications it would of interest to proceed from 1D to 2D. It is known that in two spatial dimensions the tangent discretization of $\sigma_x df/dx+\sigma_y df/dy$ still allows for a reformulation of ${\cal H}\psi=E\psi$ as a generalized eigenvalue problem \cite{Pac21}:
\begin{align}
&{\cal Q}\psi=E{\cal P}\psi,\;\;{\cal P}=\tfrac{1}{4}(1+\cos k_x)(1+\cos k_y),\\
&{\cal Q}=\tfrac{1}{2}\sigma_x(1+\cos k_y)\sin k_x+\tfrac{1}{2}\sigma_y(1+\cos k_x)\sin k_y.\nonumber
\end{align}
Therefore, we expect that an efficient 2D tensor network representation of the 2D tangent Hamiltonian in the form of Projected Entangled Pair Operators (PEPO) \cite{Cir21} can be constructed similarly to the 1D MPO approach.

\acknowledgments

We thank P. Silvi for explaining to us how long-range hopping allows for an MPO of low bond dimension. Discussions with F. Verstraete and his group helped us to understand the correspondence between our formulation of the tensor network and that of Ref.\ \onlinecite{Hae24} (see App.\ \ref{app_gep}).

C.B. received funding from the European Research Council (Advanced Grant 832256). J.T. received funding from the National Science Centre, Poland, within the QuantERA II Programme that has received funding from the European Union's Horizon 2020 research and innovation programme under Grant Agreement Number 101017733, Project Registration Number 2021/03/Y/ST3/00191, acronym {\sc tobits}. S.P. acknowledges support from Shell Gobal Solutions BV. P.E. acknowledges the support received by the Dutch National Growth Fund (NGF), as part of the Quantum Delta NL programme, as well as the support received through the NWO-Quantum Technology programme (Grant No. NGF.1623.23.006). 

\appendix

\section{Local generalized eigenproblem allows for a scale-independent MPO}
\label{app_GEV}

The DMRG approach described in the main text works because the tangent fermion Hamiltonian, while having a highly nonlocal long-range coupling, can still be described by an MPO with a low and scale-independent bond dimension. Ref.\ \onlinecite{Pac21} attributes the ``hidden locality'' of tangent fermions to the fact that their spectrum is obtained from a \textit{local} generalized eigenproblem. Here we make the connection to the scale-independent MPO explicit.

Consider 1D lattice fermions with a dispersion relation $E(k)=P(k)/Q(k)$ such that both $P(k)$ and $Q(k)$ are polynomials of finite degree in $e^{ik}$,
\begin{equation}
P(k)=\sum_{n=0}^{N_P}p_n e^{ink},\;\;Q(k)=\sum_{n=0}^{N_Q}q_n e^{ink}.
\end{equation}
For example, the tangent dispersion $E(k)=2\tan(k/2)$ corresponds to $P(k)=2i(1-e^{ik})$, $Q(k)=1+e^{ik}$. In real space the operators $P$ and $Q$ couple sites separated by at most $N_P$ or $N_Q$ lattice spacings. The generalized eigenproblem $P\Psi=EQ\Psi$ is therefore local.

Consider first the case that
\begin{equation}
Q(k)=\prod_{n=1}^{N_Q}(\alpha_n- e^{ik})
\end{equation}
has \textit{distinct} roots $\alpha_n$. The partial fraction decomposition is
\begin{equation}
\frac{P(k)}{Q(k)}=D(k)+\sum_{n=1}^{N_Q}\frac{\beta_{n}}{\alpha_n-e^{ik}}
\end{equation}
with $D(k)$ a polynomial of degree $N_P-N_Q$ (vanishing if $N_P<N_Q$). The sum over $n$
corresponds in real space to a sum over coupling terms $t_{ij}$ with an exponential spacing dependence $\propto (1/\alpha_n)^{i-j}$ for $i>j$. So in this case of distinct roots we are guaranteed to have an exact MPO representation with scale-independent bond dimension.

The situation is slightly more complicated if $Q(k)$ has repeated roots,
\begin{equation}
Q(k)=\prod_{n=1}^L(\alpha_n-e^{ik})^{\ell_n},\;\;\sum_{n=1}^L\ell_n=N_Q.
\end{equation}
The partial fraction decomposition now reads 
\begin{equation}
\frac{P(k)}{Q(k)}=D(k)+\sum_{n=1}^{L}\sum_{m=1}^{\ell_n}\frac{\beta_{nm}}{(\alpha_n- e^{ik})^m}.
\end{equation}
A term $1/(\alpha_n-e^{ik})^m$ corresponds in real space to a coupling $t_{ij}\propto (1/\alpha_n)^{i-j}\times Z(i-j)$ that is an exponential times a polynomial $Z$ in the spacing of degree $m-1$. This is still of the form \eqref{tnmdef} that allows for a scale-independent MPO \cite{Fro10}.

\section{Jordan-Wigner transformation}
\label{app_JW}

To enable the DMRG calculation, we need to convert the fermionic operators $c_{n\sigma}$ into bosonic operators $a_{n\sigma}$ (hard-core bosons, excluding double occupancy of a state). This is achieved by the Jordan-Wigner transformation,
\begin{equation}
	\begin{split}
		&c_{n\uparrow}=F_1F_2\cdots F_{n-1}a_{n\uparrow},\\
		&c_{n\downarrow}=F_1F_2\cdots F_{n-1}F_n a_{n\downarrow},
	\end{split}
\end{equation}
with fermion parity operator
\begin{equation}
	F_i=(1-2n_{i\uparrow})(1-2n_{i\downarrow})=(-1)^{n_{i\uparrow}+n_{i\downarrow}}.
\end{equation}

The transformation does not increase the bond dimension of the MPO, instead of Eq.\ \eqref{MPOtangenthelical} one now has
\begin{subequations}
	\label{MPOtangentHS}
	\begin{align}
		&H_{\rm tangent}=2it_0[M^{(1)}M^{(2)}\cdots M^{(N)}]_{1,6},\\
		&M^{(n)}=\begin{pmatrix}
			1&a_{n\uparrow}F_n&a_{n\uparrow}^\dagger F_n&a_{n\downarrow}&a_{n\downarrow}^\dagger&(2it_0)^{-1}U_n\\
			0&-F_n&0&0&0&a_{n\uparrow}^\dagger\\
			0&0&-F_n&0&0&a_{n\uparrow}\\
			0&0&0&-F_n&0&-F_n a_{n\downarrow}^\dagger\\
			0&0&0&0&-F_n&-F_n a_{n\downarrow}\\
			0&0&0&0&0&1
		\end{pmatrix}.
	\end{align}
\end{subequations}
This is for the tangent discretization. For the sine discretization the $-F_n$ on the diagonal are replaced by 0,
\begin{subequations}
	\label{MPOsineHS}
	\begin{align}
		&H_{\rm sine}=\tfrac{1}{2}it_0[M^{(1)}M^{(2)}\cdots M^{(N)}]_{1,6},\\
		&M^{(n)}=\begin{pmatrix}
			1&a_{n\uparrow}F_n&a_{n\uparrow}^\dagger F_n&a_{n\downarrow}&a_{n\downarrow}^\dagger&(\tfrac{1}{2}it_0)^{-1}U_n\\
			0&0&0&0&0&a_{n\uparrow}^\dagger\\
			0&0&0&0&0&a_{n\uparrow}\\
			0&0&0&0&0&-F_n a_{n\downarrow}^\dagger\\
			0&0&0&0&0&-F_n a_{n\downarrow}\\
			0&0&0&0&0&1
		\end{pmatrix}.
	\end{align}
\end{subequations}

\section{Periodic boundary condition for MPO with sine discretization}
\label{app_PBC_sine}

The sine discretization \eqref{tnmdefsine} requires an additional hopping term between sites 1 and $N$. The modified MPO has bond dimension 10,
\begin{subequations}
	\label{MPOsineHSperiodic}
	\begin{equation}
		H_{\rm sine}=\tfrac{1}{2}it_0[\tilde M^{(1)}\tilde  M^{(2)}\cdots \tilde M^{(N)}]_{1,6},
	\end{equation}
	\begin{equation}
		\tilde{M}^{(n)}=\begin{pmatrix}
			M^{(n)}&\delta_{n,1}W^{(n)}\\
			\delta_{n,N}W^{(n)}&(1-\delta_{n,1}-\delta_{n,N})W^{(n)}
		\end{pmatrix},
	\end{equation}
	with $M^{(n)}$ as in Eq.\ \eqref{MPOsineHS} and
	\begin{equation}
		W^{(1)}=\begin{pmatrix}
			a_{1\uparrow}F_1&a^\dagger_{1\uparrow}F_1&a_{1\downarrow}&a_{1\downarrow}^\dagger\\
			0&0&0&0\\
			0&0&0&0\\
			0&0&0&0\\
			0&0&0&0\\
			0&0&0&0
		\end{pmatrix},
	\end{equation}
	\begin{equation}
		W^{(1<n<N)}=\begin{pmatrix}
			F_n&0&0&0\\
			0&F_n&0&0\\
			0&0&F_n&0\\
			0&0&0&F_n
		\end{pmatrix},
	\end{equation}
	\begin{equation}
		W^{(N)}=\begin{pmatrix}
			0&0&0&0&0&a_{N\uparrow}^\dagger\\
			0&0&0&0&0&a_{N\uparrow}\\
			0&0&0&0&0&-F_Na^\dagger_{N\downarrow}\\
			0&0&0&0&0&-F_Na_{N\downarrow}
		\end{pmatrix}
	\end{equation}
\end{subequations}

\section{Convergence of the DMRG calculations}
\label{app_DMRG}

The tensor network formulation of the Luttinger liquid on an $N$-site chain is based on two matrix-product representations: of the operator $H$ (MPO) and of the state $\Psi$ (MPS). The MPO is exact, in terms of an $N$-fold product of $6\times 6$ matrices of creation and annihilation operators. 

The MPS is approximate: defined on $N$ sites with physical dimensional $d$, it is an $N$-fold product of $\chi \times\chi \times d$ tensors that introduces an error of order $N\sum_{n>\chi}\lambda_n^2$, with $1\geq\lambda_1\geq\lambda_2\geq\cdots\geq\lambda_{d^{N/2}}\geq 0$ the coefficients in the Schmidt decomposition of $\Psi\in{\cal H}_{1}\otimes{\cal H}_{2}$ (describing the entanglement between the first and second half of the chain, with Hilbert spaces ${\cal H}_1$ and ${\cal H}_2$) \cite{Ver06}.

\begin{figure}[tb]
\centerline{\includegraphics[width=1\linewidth]{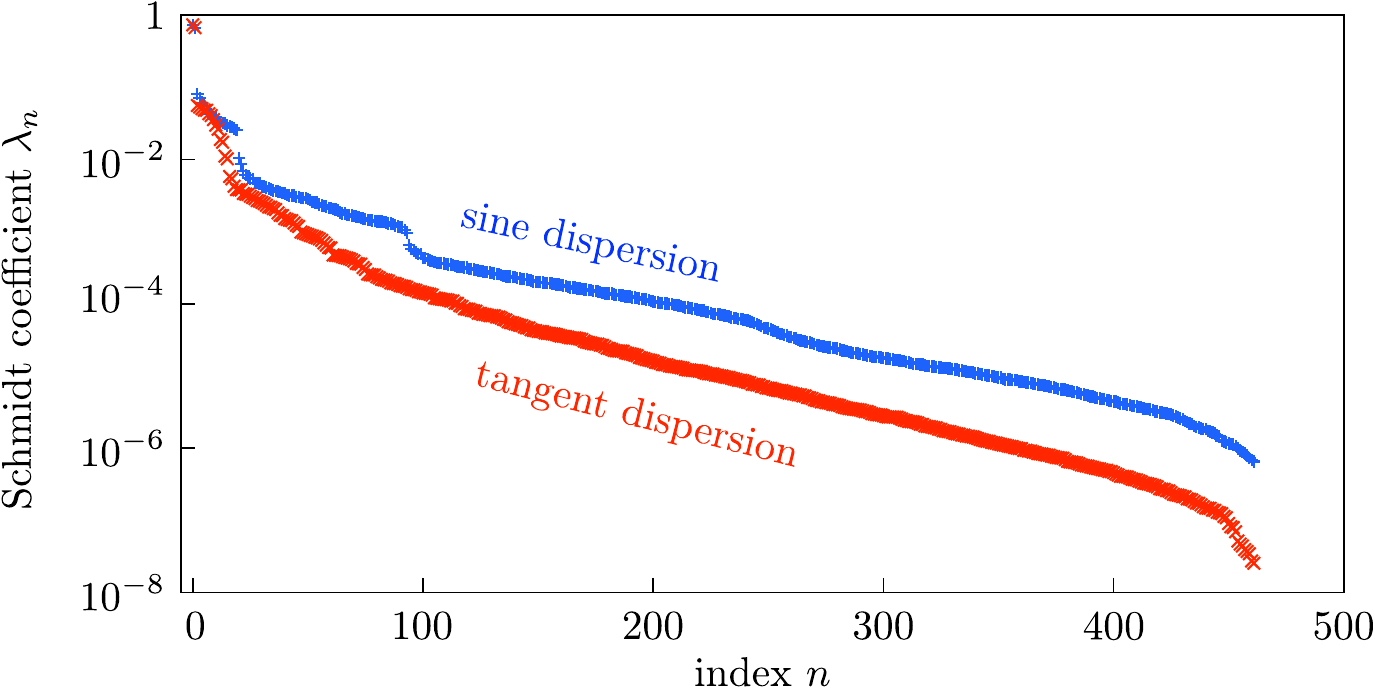}}
\caption{Log-linear plot of the Schmidt coefficients $\lambda_n$ of the ground state wave function of the Luttinger liquid Hamiltonian \eqref{HHubbard} ($U=t_0$, $N=11$ partitioned into $N_1=5$ and $N_2=6$), for the sine and tangent dispersions \eqref{tnmdefsine} and \eqref{tnmdeftangent}. The exponential decay allows for an MPS with bond dimension $\chi\ll 4^{N/2}$.
}
\label{fig_Schmidt}
\end{figure}

\begin{figure}[tb]
\centerline{\includegraphics[width=0.9\linewidth]{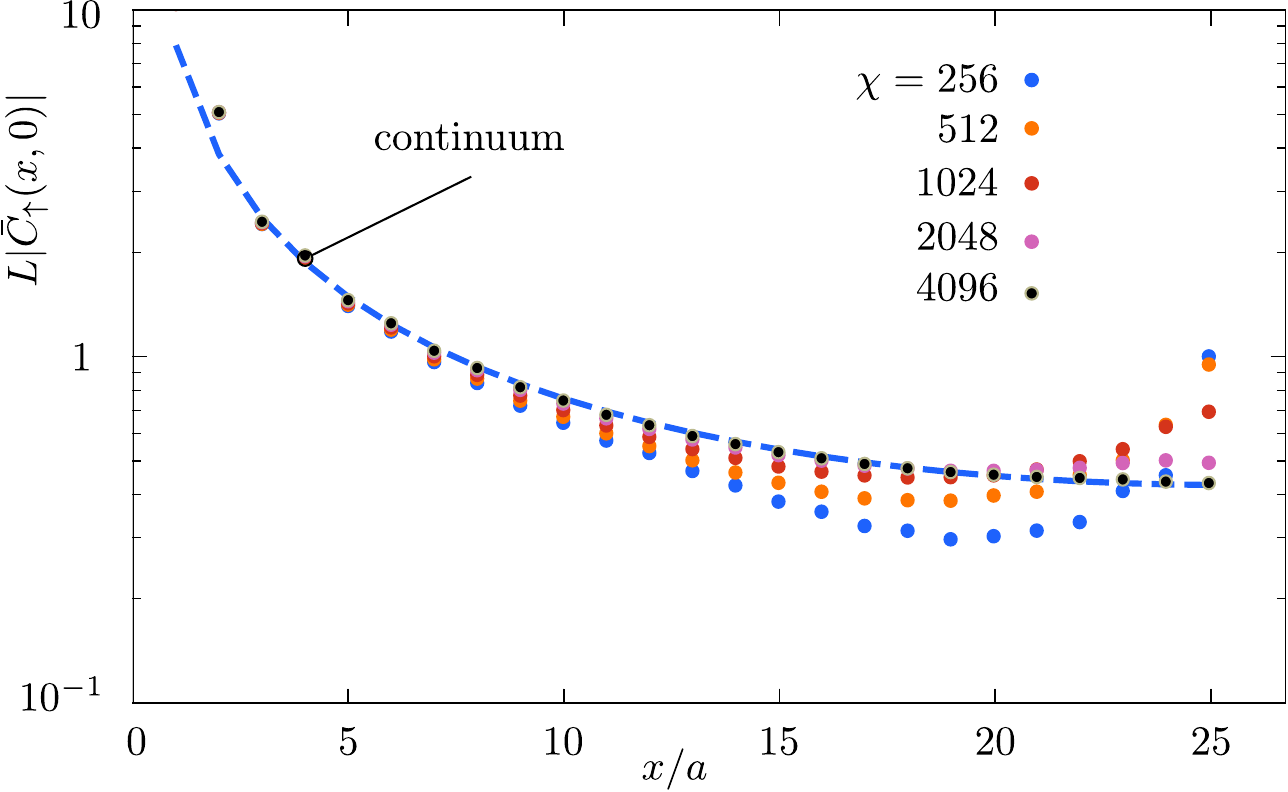}}
\caption{Dependence of the propagator on the bond dimension $\chi$ of the MPS in the tangent fermion Luttinger liquid ($L/a=51$, $\kappa=0.3$). The data in Fig.\ \ref{fig_Cplot} (black crosses) corresponds to $\chi=4096$.
}
\label{fig_Cplot_conv}
\end{figure}

The MPS is efficient at bond dimension $\chi\ll d^{N/2}$ if the Schmidt coefficients $\lambda_n$ decrease exponentially with $n$. In Fig.\ \ref{fig_Schmidt} we check this for both the sine and tangent dispersions. In Fig.\ \ref{fig_Cplot_conv} we show the convergence of the DMRG calculation with increasing $\chi$. We conclude that $\chi=4^{6} \ll 4^{25}$ (in our case $d=4$) is sufficient for the results to converge to the expected behavior.

\section{Bosonization results with finite-size effects}
\label{app_bosonization}

The power law correlators \eqref{powerlaw} follow from bosonization of the helical Luttinger liquid in the limit of an infinite system \cite{Gia03}. To reliably compare with the numerical results on a lattice of length $L$ we need to include finite size effects \cite{vanDelft98}. In Ref.\ \onlinecite{Zak24} such a calculation was reported for the grand canonical ensemble (fixed chemical potential) at finite temperature, appropriate for quantum Monte Carlo. For the DMRG calculations we need the results at zero temperature in the canonical ensemble (fixed particle number ${\cal N}={\cal N}_\uparrow+{\cal N}_\downarrow$).

The Hamiltonian of a helical Luttinger liquid with Hubbard interaction on a ring of length $L$ (periodic boundary conditions) is given by
	\begin{align}
		H ={}& \int \limits_{-L/2}^{L/2} d x \biggl(v\psi^\dagger_{\uparrow}(x)p_x\psi_{\uparrow}(x)-v\psi^\dagger_{\downarrow}(x)p_x\psi_{\downarrow}(x)\nonumber\\
		{}&  + Ua\rho_{\uparrow}(x)\rho_{\downarrow}(x) \biggr).\label{Happdef}
	\end{align}
The density $\rho_{\sigma}= \,:\!\psi^\dagger_{\sigma}\psi_{\sigma}^{\vphantom{\dagger}}\!:$ is normal ordered --- the Fermi sea of a half-filled band ($N$ particles) is subtracted. 

The bosonization results in the canonical ensemble at zero temperature are \cite{Zak24}
\begin{subequations}
\begin{align}
&C_\sigma(x,0)=\frac{\sigma e^{i\pi(2 {\cal N}_\sigma-N)x/L}}{2\pi ia_{*}|(L/\pi a_{*})\sin(\pi x/L)|^{(1/2)(K+1/K)}},\\
	&R(x,0)
	= \frac{ \cos\bigl(2\pi({\cal N}-N)x/L\bigr)}{2(2\pi a_{*})^2|(L/\pi a_{*})\sin(\pi x/L)|^{2K}},\\
&	K=\sqrt{(1-\kappa)/(1 +\kappa)},\;\;\kappa=\frac{U}{2\pi t_0}\in(-1,1).
\end{align}
\end{subequations}
The hopping energy is $t_0$ and $a_*$ is a short-distance (UV) regularization constant. For the comparison with a lattice calculation we identify $t_0=\hbar v/a$ and take $L/a=N$ an odd integer. The half-filled band corresponds to ${\cal N}_\sigma=(N+\sigma)/2$. To relate the lattice constant $a$ to the continuum regularization constant $a_*$ we argue as follows.

In the continuum theory \cite{vanDelft98} large momentum transfers $q$ are cut-off by the substitution
\begin{equation}
c^{\dagger}_{\sigma, q/2}c^{\vphantom{\dagger}}_{-q/2}\mapsto e^{-qa_*/2}c^{\dagger}_{\sigma, q/2}c^{\vphantom{\dagger}}_{-q/2}.
\end{equation}
On the lattice the averaging \eqref{barcdef} takes care of the UV regularization, 
\begin{equation}
\begin{split}
&c^{\dagger}_{\sigma, q/2}c^{\vphantom{\dagger}}_{-q/2}\mapsto\bar{c}^\dagger_{\sigma, q/2} \bar{c}_{\sigma, -q/2}^{\vphantom{\dagger}}=f(q) c^\dagger_{\sigma, q/2}c_{\sigma, -q/2}^{\vphantom{\dagger}},\\
&f(q)=\tfrac{1}{4}(1+e^{-iqa/2})^2,\;\;|f(q)|= \cos^2( q a / 4).
\end{split}
\end{equation}
We fix the ratio $a/a_*$ by equating the integrated weight factors,
\begin{equation}
\int_0^{2\pi/a}e^{-qa_*/2}\,dq=\int_0^{2\pi/a}|f(q)|\,dq\Rightarrow a/a_\ast\approx 2.
\end{equation}
 The resulting correlators are plotted in Figs.\ \ref{fig_Cplot} and \ref{fig_Rxplot}.
 
 \section{Alternative tensor network representation of Ref.\ \onlinecite{Hae24}}
 \label{app_gep}
 
 An alternative tensor network representation of the problem has been developed in Ref.\ \onlinecite{Hae24}, starting from the transformations
 \begin{equation}
\bm{a}=D^\dagger \bm{c},\;\;\bm{b}=D^{-1}\bm{c},\;\;D_{nm} = \tfrac{1}{2}(\delta_{n,m}+\delta_{n,m-1}),
\end{equation}
of the free fermion operators ${c}_n$. These are not canonical transformations, as a consequence the commutation relations of the $a$- and $b$-operators are nontrivial:
\begin{equation}
 \begin{split}
 &\{a_n, a^\dagger_m \} =(D^\dagger D)_{nm},\; \{b_n, b^\dagger_m \} = (D^\dagger D)^{-1}_{nm},\\
 &\{c_n,c^\dagger_m\}=\delta_{nm},\;\;\{b_n, a^\dagger_m \} = \delta_{nm}.
 \end{split}
\end{equation}

The corresponding $N$-fermion bases in Fock space are
\begin{equation}
\begin{split}
 	&	|\psi \rangle = \sum_{n_i=0,1}\psi^{\alpha}_{n_1,...,n_N} |n_1,...,n_N\rangle_\alpha,\\
	& |n_1,...,n_N\rangle_\alpha = (\alpha^\dagger_{1})^{n_1}...(\alpha^\dagger_{N})^{n_N},
 \end{split}
 \end{equation}
 with $\alpha \in \{a,b,c\}$. Only the $c$-basis is orthonormal, the two other bases produce non-diagonal norm matrices $\tilde{N}$,
 \begin{align}
	_a\langle m_1,...,m_N|n_1,...,n_N\rangle_a = \tilde N^{m_1,...,m_N}_{n_1,...,n_N},\nonumber\\
	_b\langle m_1,...,m_N|n_1,...,n_N\rangle_b = (\tilde N^{-1})^{m_1,...,m_N}_{n_1,...,n_N},\nonumber\\
	_c\langle m_1,...,m_N|n_1,...,n_N\rangle_c = \delta_{m_1n_1}\cdots\delta_{m_Nn_N}.
 \end{align}
The $a$ and $b$ bases are bi-orthogonal,
 \begin{equation}
	_a\langle m_1,...,m_N|n_1,...,n_N\rangle_b =\delta_{m_1n_1}\cdots\delta_{m_Nn_N}.
\end{equation}

The motivation for these transformations is that the tangent fermion Hamiltonian becomes local in terms of the $b$-operators, 
 \begin{subequations}
 \begin{align}
 	H_{\rm tangent}&=2it_0\sum_{n>m=1}^N (-1)^{n-m}\bigl(c_n^\dagger c_m^{\vphantom{\dagger}}-c_m^\dagger c_n^{\vphantom{\dagger}}\bigr)\\
 	&=\frac{t_0}{2i}\sum_{n=1}^N\bigl(b_{n+1}^\dagger b_{n}^{\vphantom{\dagger}}-b_n^\dagger b_{n+1}^{\vphantom{\dagger}}\bigr).
 \end{align}
\end{subequations}
Matrix elements of $H_{\rm tangent}$ in the $a$-basis, orthogonal to the $b$-basis, can therefore be evaluated efficiently.

The key step of Ref.\ \onlinecite{Hae24} is to derive a scale-independent MPO representation of the norm matrix $\tilde{N}$ in the $a$-basis. We have followed a different route, we stay with the orthonormal $c$-basis and a nonlocal Hamiltonian, but we have found that it does not stand in the way of a scale-independent MPO representation.


\begin{thebibliography}{99}
\bibitem{note1} The tangent discretization \eqref{tnmdeftangent} corresponds to Eq.\ \eqref{tnmdef} with $p=0$ (purely exponential distance dependence). More generally, one can take $p=1,2,\ldots$ (polynomial-times-exponential distance dependence),  in which case the dispersion contains terms $\propto \tan(k/2)(1+\cos k)^{-p}$. The MPO representation of the discretized Hamiltonian remains scale independent, but the bond dimension is larger than for $p=0$.
\bibitem{note0} The Fourier series $\tan(k/2)=-2\sum_{n=1}^\infty(-1)^n\sin nk$ implied by the identification \eqref{tnmdeftangent} should be understood in the sense of a distribution: $\int_{0}^\pi \tan(k/2)f(k)\,dk=-2\sum_{n=1}^\infty (-1)^n \int_{0}^\pi f(k)\sin nk\,dk$, with $f(k)$ a  test function that vanishes at $\pi$.
\bibitem{note2} As a consistency check, we note that in real space the fact that an odd (even) number $N$ of lattice sites corresponds to periodic (antiperiodic) boundary conditions, ensures in momentum space that the discrete wave numbers avoid the pole of the tangent dispersion relation at the Brillouin zone boundary.
\end{thebibliography}
\end{document}